\documentclass[conference]{IEEEtran}
\usepackage[normalem]{ulem}

\usepackage{ifpdf}

\usepackage{cite}

\ifCLASSINFOpdf
\else
\fi
\usepackage{amsmath}
\usepackage{algorithmic}

\usepackage{array}

 \usepackage[caption=false,font=normalsize,labelfont=sf,textfont=sf]{subfig}
\usepackage{fixltx2e}

\usepackage{stfloats}
\usepackage{url}

\usepackage{xcolor}
\usepackage{multirow}
\definecolor{new_color}{RGB}{0, 0, 255}
\definecolor{kelly_color}{RGB}{255, 0, 0}

\begin{document}
%
\title{Throwaway Accounts and Moderation on Reddit}

\author{\IEEEauthorblockN{Cheng Guo}
\IEEEauthorblockA{Clemson University\\
chengg@clemson.edu}
\and
\IEEEauthorblockN{Kelly Caine}
\IEEEauthorblockA{Clemson University\\
caine@clemson.edu}}
	

%


\IEEEoverridecommandlockouts
\makeatletter\def\@IEEEpubidpullup{6.5\baselineskip}\makeatother
\IEEEpubid{\parbox{\columnwidth}{
		Symposium on Usable Security and Privacy (USEC) 2025 \\
		24 February 2025, San Diego, CA, USA \\
		ISBN 979-8-9919276-5-9 \\
		https://dx.doi.org/10.14722/usec.2025.23031 \\
		www.ndss-symposium.org, https://www.usablesecurity.net/USEC/
}
\hspace{\columnsep}\makebox[\columnwidth]{}}

\maketitle

\begin{abstract}
Social media platforms (SMPs) facilitate information sharing across varying levels of sensitivity. A crucial design decision for SMP administrators is the platform's identity policy, with some opting for real-name systems while others allow anonymous participation. Content moderation on these platforms is conducted by both humans and automated bots. This paper examines the relationship between anonymity, specifically through the use of ``throwaway'' accounts, and the extent and nature of content moderation on Reddit. Our findings indicate that content originating from anonymous throwaway accounts is more likely to violate rules on Reddit. Thus, they are more likely to be removed by moderation than standard pseudonymous accounts. However, the moderation actions applied to throwaway accounts are consistent with those applied to ordinary accounts, suggesting that the use of anonymous accounts does not necessarily necessitate increased human moderation. We conclude by discussing the implications of these findings for identity policies and content moderation strategies on SMPs.
\end{abstract}


%
\IEEEpeerreviewmaketitle

\section{Introduction}
 Social Media Platforms (SMPs), including Reddit, facilitate information sharing and discussion across a wide range of topic sensitivities. Identity policies, which determine whether users must participate using real names or can remain anonymous, vary across different platforms. Facebook, for example, enforces a strict real-name policy\footnote{https://www.facebook.com/help/112146705538576}, while others, such as Reddit, allow users to register with any desired username. These usernames, although not real names, provide a degree of identity continuity and are considered pseudonyms. Certain SMPs, like 4chan, offer complete anonymity through system-generated placeholders for screen names and avatars, effectively delinking user activities from their accounts. SMPs like Reddit employ a distinctive identity model. Although Reddit does not explicitly offer anonymity, users have developed a workaround through the use of ``throwaway'' accounts~\cite{leavitt2015throwaway}. This practice aligns with broader internet user behavior aimed at masking online activities~\cite{rainie2013anonymity}. On Reddit, throwaway accounts are those that Reddit users create intentionally for temporary use and are unlinkable to their primary accounts, aiming for short-term anonymity. In this paper, we adopt the terms ``throwaway'' and ``identified''~\cite{andalibi2016understanding,andalibi2018social} to differentiate between these temporary anonymous accounts and standard pseudonymous accounts on Reddit.

The use of anonymity on SMPs has been linked to a potential lack of consequences for irresponsible behavior~\cite{kang2013people}. Irresponsible behavior is prevalent on SMPs and can manifest even in ordinary individuals under certain circumstances~\cite{cheng2017anyone}. Consequently, Reddit, like other SMPs, requires effective regulations for successful operation~\cite{kraut2012building}. Moderation, as a form of regulation, is now employed on most SMPs, including Reddit, to address irresponsible behavior such as trolling and online harassment~\cite{herring2002searching}. Subreddits~(``communities dedicated to specific topics, where redditors can post content and interact with one another''\footnote{https://support.reddithelp.com/hc/articles/204533569}) on Reddit experience varying degrees of moderation~\cite{andalibi2018social}, which can result in content removal~\cite{grimmelmann2015virtues}. However, the relationship between content moderation and anonymous identities (i.e., throwaway accounts) remains unclear. Previous research indicates that anonymous content receives more downvotes on social Q\&A sites like Yahoo Answers~\cite{guo2020anonymity}, potentially suggesting a higher likelihood of removal due to downvotes being a predictor of low-quality content~\cite{ponzanelli2014improving}. However, Guo and Caine found that anonymous and non-anonymous answers on Q\&A sites exhibited similar quality levels~\cite{guo2021anonymity}, but their study was limited by the content moderation mechanisms of the platforms. Researching raw, unmoderated data from SMPs is highly challenging and, in most cases, impractical or unfeasible. Almost all SMPs implement moderation mechanisms, meaning that low-quality or inappropriate content may already have been moderated and removed before researchers can access the data. Furthermore, moderation logs and histories are typically not publicly available for researchers to analyze. As a result, a portion of the data is often missing, which could introduce bias in studies involving SMPs. Similar pre-collection moderation may have occurred on Reddit as well~\cite{andalibi2018social}. Given the potential for anonymity to influence content removal by moderation and the ambiguity surrounding the relationship between identity and moderation outcomes, we pose the following research question:

\begin{itemize}
\item \textit{RQ1}: What is the difference in the level of removal by moderation between throwaway accounts and identified accounts on Reddit?
\end{itemize}

The question of who bears the responsibility for content moderation remains complex. Moderating content on SMPs, particularly at scale, is challenging~\cite{kiene2019technological}. Consequently, alongside the traditional approach of human moderation, machine moderation (e.g., moderation bots) is increasingly used on platforms such as Reddit~\cite{jhaver2019human}. However, human moderation incurs significant costs, both in terms of time and psychological toll~\cite{humancost:1,humancost:2}. It remains unclear whether the use of anonymity further exacerbates these costs by requiring additional human effort to moderate and remove content versus just utilizing moderation bots. Therefore, we pose a second research question:

\begin{itemize}
\item \textit{RQ2}:
Does the of throwaway accounts on Reddit increase the likelihood of content removal by human moderators compared to removal by moderation bots?
\end{itemize}

In this study, we investigate the prevalence of throwaway accounts on Reddit, specifically focusing on instances where content is removed by either moderation bots or human moderators. We utilize data provided by publicmodlogs\footnote{https://www.reddit.com/user/publicmodlogs}, compiling a dataset encompassing all moderation logs, removed content, and unremoved content from 340 subreddits over a three-month period. A post-hoc observational analysis of this dataset aims to enhance our understanding of the relationship between throwaway accounts and content moderation practices on Reddit.

Reddit is a distinctive platform where community norms promote the use of throwaway accounts for temporary anonymity, while most content is contributed by pseudonymous accounts. It also has a comprehensive moderation system that involves both bots and human moderators, with some moderation logs being publicly accessible. These features make Reddit an ideal platform for studying throwaway accounts and moderation, providing valuable insights into the relationship between anonymity and content moderation on SMPs. 

Our findings demonstrate that: (1) content originating from throwaway accounts on Reddit is more likely to violate rules and has a higher likelihood of being removed by moderation; (2) moderators assess anonymous content equitably, as it is treated similarly to content posted by identified accounts; and (3) while moderation on Reddit still heavily relies on human moderators, the use of anonymous accounts does not appear to increase the demand for human intervention in content removal.

\section{Background and Related Work}
Although providing a comprehensive review of the extensive and growing literature on identity and content moderation is outside the scope of this paper, in this section, we review key theoretical and empirical work in these areas that are most relevant to our work.

\subsection{Anonymity and Throwaway Accounts on Reddit}
\subsubsection{Types of identity on SMPs}
People use different types of identity on SMPs, ranging from real names to anonymity. SMPs like Facebook have a strict real-name policy in their terms and conditions that require users to use their authentic names. Although this real-name policy is controversial and some users use fake names and/or identities to get around it~\cite{haimson2016constructing}, Facebook actively detects fake accounts using algorithms and encourages its users to report them~\cite{facebook:fake}. On the other hand, many other SMPs, including Reddit, rely on a pseudonymity model where users create a screen name when they access the sites. Still, other SMPs (e.g., 4chan) use an anonymity identity policy. In this model, all posts are anonymous in the  sense that they are not associated with a user-created screen name. A common limitation across these models is the lack of flexibility in identity choice. For instance, Facebook exclusively offers real names, preventing pseudonymous or anonymous participation. This inflexibility may not align with user preferences, as individuals may desire different levels of identity disclosure depending on the context or content sensitivity~\cite{peddinti2014cloak}.

\subsubsection{The community norm of using throwaway accounts on Reddit}
Reddit does not provide a formal anonymity feature that allows users to use it anonymously. However, we observe that users still have the desire to engage on Reddit anonymously in certain situations. The community norm, in this case, is for Reddit users to engage in a workaround approach where they use a throwaway account. Throwaway accounts on Reddit are temporary user accounts used only for a specific time and purpose. Reddit users use throwaway accounts to share secrets or discuss sensitive information without being identified and with the ability to walk away from further discussion~\cite{gagnon2013disinhibition}. As a result, many throwaway accounts are only used once~\cite{gagnon2013disinhibition}, and almost all (96.3\%) throwaway accounts on Reddit are used three or fewer times~\cite{leavitt2015throwaway}.

\subsubsection{Throwaway accounts on Reddit are anonymous}
Marx argues that ``to be fully anonymous means that a person cannot be identified according to any of the seven dimensions of identity knowledge.''~\cite[pg. 100]{marx1999s} Those dimensions are: 1) legal name 2) locatability 3) pseudonyms that can be linked to legal name and/or locatability\---literally a form of pseudonymity 4) pseudonyms that cannot be linked to other forms of identity knowledge\---the equivalent of ``real'' anonymity (except that the name chosen may hint at some aspects of ``real'' identity, as with undercover agents encouraged to take names close to their own) 5) pattern knowledge 6) social categorization 7) symbols of eligibility/non-eligibility. Marx further argues that we have true anonymity only when ``no aspects of identity are available (being uncollected, altered, or severed).''~\cite[pg. 104]{marx2019windows} 

Absolute anonymity on the Internet is difficult to achieve, even for experts, and some argue, may not even be possible~\cite{marx2004internet}. Although throwaway accounts on Reddit do not perfectly fit the dimensions of anonymity Marx describes (people may still self-disclose in their posts/comments when using throwaway accounts), prior work~\cite{leavitt2015throwaway,gagnon2013disinhibition,urbanski2013upvoting} suggests that throwaway accounts can serve as proxies for true anonymity, as they provide an acceptable level of anonymity. Thus, following prior studies (e.g., ~\cite{andalibi2016understanding,andalibi2018social,pavalanathan2015identity},  we use the terms ``identified'' and ``anonymous'' to distinguish between ordinary accounts and throwaway accounts on Reddit.

\subsubsection{Other works studying throwaway accounts on Reddit}
Our work is not the only work that investigates throwaway accounts on Reddit. For example, other work finds that factors including the perception of anonymity and gender affect the ways people use throwaway accounts on Reddit~\cite{leavitt2015throwaway}. Within subreddits that focus on marginalized groups (e.g., Asian Americans and Pacific Islanders focused subreddits), conservative people are more likely to post using throwaway accounts~\cite{dosono2017exploring}. Using an anonymous identity via a throwaway account can bring certain benefits, such as enabling users to express views and thoughts more freely, especially about sensitive topics such as mental health~\cite{de2014mental}. Furthermore, the use of throwaway accounts does not hinder the quality of social support users receive from other Redditors~\cite{de2014mental}. Similarly, throwaway accounts allow users to share intimate content but are not associated with aggressive or unsupportive comments~\cite{andalibi2018social}. On Reddit, comments posted using throwaway accounts are more likely to receive a response, receive longer responses, and receive responses with higher Karma scores (users get Karma if their content gets upvoted) than those posted using identified accounts~\cite{ammari2019self}.

Besides all these benefits, however, online anonymity also has drawbacks, such as resulting in a lack of consequences for deviant behavior~\cite{johnson2000anonymity}. Deviant behavior such as trolling is common on SMPs such as Reddit~\cite{krappitz2012troll}. According to the social identity theory of deindividuation~\cite{reicher1995social}, people in groups (e.g., an SMP such as Reddit) tend to lose selfhood and thus lose self-control over behavior. In this theory, anonymity could lead to this loss of control. This loss of control could then lead to deviant behavior, which may violate the community norms that are likely to be targets of moderation on Reddit, and are likely to be removed by moderation ~\cite{chandrasekharan2018internet}. While anonymous editors on Wikipedia are more prone to violating policies related to edit warring, particularly on discussion pages~\cite{tran2020anonymity}, Reddit users engage with the platform for a wider range of activities beyond information sharing and editing~\cite{bogers2014social,moore2017redditors}. So far, we don't yet know if the use of throwaway accounts on Reddit will have any relationship with content removal by moderation. Thus, we extend prior research in this area by asking RQ1.

\subsection{Content Moderation on Reddit}\label{sec:moderation}
\subsubsection{Two types of moderation on SMPs}
On SMPs, content moderation fits into two categories: distributed moderation and centralized moderation~\cite{chandrasekharan2018internet}. Distributed moderation is a moderation mechanism that distributes moderation tasks to site users. For example, many social Q\&A sites use a voting system that users can upvote and downvote each others' Q\&As~\cite{wang2013wisdom,li2015answer}. Then the voting outcome ``determines how prominently any content is displayed on the site'', which ``allows the community to collectively decide its threshold for what content is acceptable and which issues need to be articulated and discussed.''~\cite{chandrasekharan2018internet}

\subsubsection{Centralized moderation}
Unlike distributed moderation which allows users to conduct moderation activities, centralized moderation uses moderators to handle moderation tasks. For example, SMPs like YouTube actively recruit human moderators to do moderation tasks on their sites~\cite{kraut2012building}. Reddit adopts both distributed moderation and centralized moderation approaches. On the one hand, users can upvote and downvote other users' posts/comments. This voting system influences the visibility and prominence of content on the platform's interface. On the other hand, there are a certain number of users that voluntarily serve as human moderators for each subreddit. On Reddit, only centralized moderation will lead to content removal. The distributed moderation (users' votes) will not directly lead to content removal. Since our research questions are around content moderation that leads to content removal, in this work, we focus on the latter\---centralized moderation on Reddit.

\subsubsection{Centralized moderation on Reddit}
Besides human moderators, Reddit also enables moderation bots\footnote{https://www.reddit.com/r/TheoryOfReddit/wiki/bots/}, which are popular as a class of automated moderation tools supporting Reddit moderators~\cite{kiel2017could}. Among these moderation bots, the most widely used one is named ``AutoModerator''~\cite{AutoModerator}. AutoModerator is an open-sourced site-wide tool set up by human moderators to facilitate the moderation tasks, especially repetitive ones~\cite{AutoModerator}. Human moderators can also develop bots using Reddit's API to perform similar and/or customized tasks. For both moderation bots and human moderators, there is a variety of list of moderation actions they could do on Reddit\footnote{https://support.reddithelp.com/hc/articles/15484284113172}. These moderation actions could lead to different outcomes. Some of them might lead to the removal of user content. TeBlunthuis, Hill, and Halfaker found out that on SMPs like Wikipedia, human moderators can focus on social signals but overlook the actual misbehavior~\cite{teblunthuis2021effects}. They also found that utilizing algorithmic flagging (which is similar to the moderation bots on Reddit) can reduce such bias and increase moderation fairness. Likewise, on Reddit, human moderators can focus on social cues like the usernames rather than the actual content users post. It is unclear that if the content posted by throwaway accounts is more likely to be removed by moderation by human moderators than by moderation bots. Thus, we ask RQ2.
\section{Methods}
\subsection{Dataset}
Publicmodlogs is an account on Reddit that makes all moderation logs of certain subreddits public. Moderators of any subreddit can voluntarily invite publicmodlogs as one of their subreddit's moderators. Publicmodlogs then posts all the moderation logs of the subreddit automatically to its account. Thus, the moderation logs of the subreddit become publicly available. 
When we collected our data in October 2019, 340 subreddits voluntarily participated in publicmodlogs. To our knowledge, our dataset has the most diverse subreddits compared to all other studies that use data from publicmodlogs. Juneja et.al's data was collected in 2018 and contained 204 subreddits~\cite{juneja2020through}. Li et.al's data was collected between 2020 and 2021, and contained 84 subreddits~\cite{li2022all,li2022measuring}. The topics of these subreddits vary from highly sensitive to less sensitive content. 
Using throwaway accounts becomes a community norm in sensitive subreddits~\cite{van2013faceless}. By using publicmodlogs, we hope to shed light on a more comprehensive understanding of the use of anonymity and content moderation on SMPs.

\subsubsection{Moderated Content Collection}\label{section:moderated}
To answer both of our research questions, we collected a subset of moderated and removed content. Using publicmodlogs and Reddit's API, we collected the most recent 500 moderation logs, which contained moderation actions for both posts and comments of all 340 subreddits in mid-October of 2019. We wrote a Python JSON parser to parse the JSON objects returned by Reddit's API. Using this approach, we were able to collect the following information about each moderation log: the moderation ID, date and time the moderation happened, the moderation action, the name of moderators, the name of users whose post/comment was moderated, and the content of each post/comment. We collected 36,514 moderation logs in total. The dates of these logs ranged from mid-July to mid-October of 2019.

\subsubsection{Unremoved Content Collection}\label{section:unmoderated}
We additionally collected a subset of unremoved content. Using the Pushshift API\footnote{https://pushshift.io}, we collected all the unremoved posts and comments of all 340 subreddits within the period of those moderated logs (mid-July, 2019 to mid-October, 2019). We wrote a Python JSON parser to parse the JSON objects returned by the Pushshift API. Pushshift API stores comments and posts from Reddit in real-time as long as they are not removed by moderation bots (e.g., AutoModerator, the moderation of which occurs immediately without delay\footnote{https://support.reddithelp.com/hc/articles/15484574206484}). This means those posts and comments that human moderators eventually removed are also included in Pushshift. For this content, the original text returned by the API will be replaced with [``removed'']~\cite{chandrasekharan2018internet}. Since we already collected those comments using publicmodlogs when we collected moderated content, we removed these duplicates from our dataset in this step. Since not all moderation actions lead to content removal, we also filtered the duplicates in this step. Using this approach, we were able to collect the following information on the unremoved content: unremoved post/comments, the author of each post/comment, date and time, and the content of each post/comment. As a result, we collected 527,763 unremoved posts and comments in total.

\begin{table*}[!htb]
    \centering
    \begin{tabular}{l|l|r}
        Sensitivity & Example subreddits & N (\%)\\
        \hline
        Not at all sensitive & r/bonehurtingjuice, r/ethereum, r/btc, r/ElderScrolls & 203 (59.7\%)\\
        Slightly sensitive & r/knives, r/YangForPresidentHQ, r/torrentlinks & 45 (13.2\%)\\
        Moderately sensitive & r/pussypassdenied/, r/collapse, r/TheseFuckingAccounts & 41 (12.1\%)\\
        Very sensitive & r/conspiracy, r/RBI, r/MeanJokes, r/liberalgunowners & 23 (6.8\%)\\
        Extremely sensitive & r/hotwife, r/tanlines, r/WatchRedditDie, r/horny & 28 (8.2\%)
    \end{tabular}
    \caption{The distribution and the examples of the sensitivity of all subreddits that participate in publicmodlogs.}
    \label{tab:sensitivity}
\end{table*}




\subsection{Measuring Sensitivity}
On SMPs such as Q\&A sites, the use of the anonymity feature is associated with the level of sensitivity of the topics~\cite{guo2021anonymity}. Thus, on Reddit, the use of throwaway accounts may be associated with different topics of the subreddits. To understand the different levels of sensitivity of each subreddit, two research assistants voluntarily served as human coders to manually rate the sensitivity of each subreddit. Following~\cite{guo2021anonymity}, human raters were trained with a broad definition of sensitivity that considers anything that could lead to discrimination as sensitive. A broad definition of sensitivity like this is a ``\textit{major advantage}'' that it allows for the ``\textit{inclusion of topics that ordinarily might not be thought of as `sensitive'}~''~\cite[p.~511]{lee1990problems}. Human raters first reviewed the content policy of Reddit. Then, they individually rated 50 randomly selected subreddits from the set of 340 in a five-point Likert scale (1-not at all sensitive, 2-slightly sensitive, 3-moderately sensitive, 4-very sensitive, 5-extremely sensitive). Afterward, they met with the the authors of this paper and discussed ratings to resolve disagreements. Then, the two raters used the annotation from these 50 subreddits as examples to rate the rest of the subreddits individually. Finally, they met with the broad research team again and discussed results from the second round and solved any remaining disagreements. Using the guidelines from work on the assessment of inter-rater reliability (IRR) that suggest using intraclass correlation (ICC) for Likert scales~\cite{hallgren2012computing}, we performed ICC on the two raters' ratings. The agreements were .702 and .786 for the first round and the second round of the rating process, both indicating good agreement~\cite{bland1997statistics}. The mean sensitivity of all subreddits is 1.91 ($\pm$ 0.07). The distribution and the examples of the sensitivity of all subreddits can be found in Table~\ref{tab:sensitivity}. 

\subsection{Measuring Activity Level}
The 340 subreddits vary in terms of user activity. Since we don't have access to Reddit's daily active user (DAU) data for subreddits, we used Reddit's API\footnote{https://www.reddit.com/dev/api/} to retrieve the number of active users online on an hourly-basis, which we used to measure the level of activity of each subreddit. Reddit's users are from all over the world, so they use Reddit in different time zones, which affects the level of activity at any one time. Therefore, we wrote a Python script to call the API to retrieve the number of active users online for each subreddit (i.e., number of users who are online and visiting the subreddit at the moment when we call the API) every hour for a one-week period. Although we don't have access to Reddit's DAU of each subreddit, we used the average number of active online users as a proxy to represent the level of activity for each subreddit. The range of the average number of hourly active online users per subreddit varies from 1.2 to 2297.9 (mean = 77.8 $\pm$ 15.6, median = 2.1). The twenty-fifth percentile is 1.6, the fiftieth  percentile is 2.1, the seventy-fifth percentile is 8.5, the ninety percentile is 84.3. Since the top quartile (75\% - 100\%) has a large variance, we used an adjusted quartile split of the average number of active online users per subreddit to create four groups: most active, very active, moderately active, and least active (See Table~\ref{tab:activity}). As shown, even though the least active category contains 50\% of the subreddits, the range is less than one.
\begin{table}[!htb]
    \centering
    \begin{tabular}{r|r|r|r}
        Level of activity &  Percentile & Hourly active users & N (\%)\\
        \hline
        Most active & 90\% - 100\% & 84.3 - 2297.9 & 32 (9.4\%)\\
        Very active & 75\% - 90\% & 8.5 - 84.3 & 54 ( 15.9\%)\\
        Moderately active & 50\% - 75\% & 2.1 - 8.5 & 84 (24.7\%)\\
        Least active & 0 - 50\% & 1.2 - 2.1 &  170 (50\%)\\
    \end{tabular}
    \caption{The level of activity split into quartiles of all subreddits that participate in publicmodlogs, using number of hourly active users as a proxy}
    \label{tab:activity}
\end{table}

\subsection{Identifying Moderation Bots}
There are two types of moderation bots on Reddit: AutoModerator and other bots built by human moderators. The screen name of an AutoModerator in a moderation log on Reddit is ``AutoModerator''. Although it is difficult to perfectly identify all moderation bots, following~\cite{jhaver2019does}'s approach, we first identify moderation bots using a list of known bot accounts (which contains ``AutoModerator'') on Reddit. Then we programmatically checked if a moderator account name contains ``bot'' in it by ignoring the case sensitivity. In addition, the first author followed up by manually checking each account to verify that the account is a moderation bot. 

\subsection{Identifying Throwaway Accounts}
Following the approach used by prior studies (i.e.,~\cite{andalibi2018social,gagnon2013disinhibition,leavitt2015throwaway,pavalanathan2015identity,ammari2019self}), we used a two-step approach to identify throwaway accounts on Reddit. First, we examined the usernames of each user programmatically to see if the username contains the word ``throwaway'' or a lexical variation of the word (e.g., *thrw*, *throwaway*, *throw*, *thraway*)~\cite{andalibi2018social}. Next, we programmatically looked for the word ``throwaway'' in post and added accounts who used statements like, “this is a throwaway account,” or “I’m using a throwaway account."~\cite{ammari2019self} In addition, the first author followed up manually to check each post to verify that ``throwaway'' was used to represent an anonymous disclosure, rather than used in the context of the post (e.g., a post containing ``throw away the trash'' would NOT indicate a throwaway account).

\subsection{Ethical Considerations}
The entire research protocol was approved by our institution's Institutional Review Board (IRB). We also adhered to Reddit's Terms of Service\footnote{https://www.redditinc.com/policies/user-agreement} and limited our data collection to publicly available information. Specifically, we collected data only from subreddits who voluntarily participate in publicmodlogs, making their moderation logs publicly accessible online. However, we acknowledge that the definition of public data is still evolving~\cite{zimmer2010but}. We endeavored to adhere to the HCI community's ethical norms for studying SMPs~\cite{fisher2010terms,fiesler2015ethics,munteanu2019sigchi,bruckman2017cscw}, such as data anonymization. All collected data is securely stored, with access restricted to the research team. After identifying throwaway accounts and moderation bots, all account names were replaced with unidentifiable labels for subsequent data processing and analysis.

\section{Results}
We first provide an overview of all the moderation logs we collected via publicmodlogs. 
Since moderation may lead to different outcomes, including content removal, we then build a logistic regression model to test the effect of type of account, the sensitivity of subreddit, level of activity of subreddit, and entry type of the content on the unremoved content vs. removed content. As moderation can be done by a moderation bot or a human moderator, afterward, we build a second logistic regression model using the removed content only to test whether content posted by throwaway accounts is more likely to be removed by human moderators or moderation bots.

For both logistic regression models we built, we tested the model for interaction effects between all predictors. The F-ratio tests suggest that adding any combination of the interaction effects did not significantly improve the model fit. Thus, we removed the interaction effects from all models. We also tested the multicollinearity of all models. All the Variance Inflation Factors (VIFs) were well below the threshold of four~\cite{o2007caution}, indicating that there was no multicollinearity issue with any of our models.

\subsection{Independent variables}\label{sec:independent}
The independent variables of both logistic regressions we conducted were following: the identity of the user account (categorical with identified accounts as the baseline); the sensitivity of the subreddit (numerical from 1-not at all sensitive to 5-extremely sensitive); the level of activity of the subreddit (numerical from 1-least active to 4-most active); the entry type of the content (categorical with posts as the baseline).



\begin{table}[!htb]
\centering
\begin{tabular}{l|rlll}
Total moderation logs & 36,514 &  &  &  \\
\hline
Removed              & 14,878 &  &  &  \\
\hspace{5mm}Throwaway            & 48     &  &  &  \\
\hspace{5mm}Identified           & 14,830  &  &  &  \\
\hline
Unremoved            & 21,636 &  &  &  \\
\hspace{5mm}Throwaway            & 58     &  &  &  \\
\hspace{5mm}Identified           & 21,578 &  &  & \\
\end{tabular}
\caption{Moderation logs collect via publicmodlogs, break down by moderation led to content removed or unremoved and content posted by throwaway accounts or identified accounts}
\label{table:logs}
\end{table}

\begin{table*}[!htb]
    \centering
    \begin{tabular}{l|rrrrr}
        & \multicolumn{5}{c}{Removed vs. Unremoved}\\
        \hline
        Factors & \textit{B(SE)} & \textit{p} & \textit{2.5\% CI} & \textit{Odds Ratio} & \textit{97.5\% CI} \\
        (Intercept)  & -0.79(0.04) & \textbf{\textless{}.001} & 0.302 & 0.330 & 0.361\\
        \begin{tabular}[c]{@{}l@{}}Throwaway accounts\\ \textit{(baseline: identified accounts)}\end{tabular} & 0.54(0.15) & \textbf{\textless{}.001} & 1.259 & 1.715 & 2.279\\
        \textit{Sensitivity} & -0.31(0.01) & \textbf{\textless{}.001} & 0.727 & 0.735 & 0.743\\
        Activity & -0.23(0.01) & \textbf{\textless{}.001} & 0.776 & 0.794 & 0.813\\
        \begin{tabular}[c]{@{}l@{}}Comments\\ \textit{(baseline: posts)}\end{tabular} & -2.04(0.02) & \textbf{\textless{}.001} & 0.126 & 0.130 & 0.134 \end{tabular}
    \caption{Effect of type of accounts, sensitivity of subreddit, level of activity of subreddit, and entry type of content on type of content (removed content vs. unremoved content ) on Reddit. Unremoved was coded as 0 and removed was coded as 1. Model fit: $\chi^2$ (4) = 17147.59, \textit{p} \textless{} .001, Nagelkerke $R^2$ = 0.14. \textit{B} is the coefficient. \textit{SE} is the standard error. \textit{CI} is the confidence interval. We use \textit{ODDS RATIO} as the effect size.}
    \label{table:regression_deletion}
\end{table*}

\subsection{Moderation logs}\label{section:moderation_actions}
We collected 36,514 moderation logs from 340 subreddits via publicmodlogs (see Table~\ref{table:logs}). The content of 106 (0.3\%) logs were posted by throwaway accounts and the rest were posted by identified accounts. The ratio of posts/comments posted by throwaway accounts versus posted by identified accounts is similar to what is reported in the literature~\cite{ammari2019self}. Among all moderation logs, 14,878 (40.7\%) of them led to content removal and 21,636 (59.3\%) did not.

In this work, we are primarily interested in the moderation logs that result in content removal. Moderation logs are created for a variety of reasons on Reddit.\footnote{https://support.reddithelp.com/hc/articles/15484543117460} For example, a moderation log is created when a moderator edits the stylesheet or the tag. A moderation log will also be created when a moderation bot or a human moderator removes inappropriate content. The moderation logs that result in content removal are: \textit{remove comment}, \textit{remove link}, \textit{spam comment}, and \textit{spam link} (``link'' here means the entire post). Table~\ref{table:moderation_full} in Appendix contains a complete list of all types of moderation logs we collected.


\subsection{Unremoved Content vs. Removed Content by Moderation}\label{section:content_deletion}
As we showed above, some moderation logs may reveal content removal and others may not. Understanding this, to explore RQ1, we constructed a logistic regression model to explore the effect of the independent variables (see Section Independent variables) on content removed by moderation vs. unremoved content (dependent variable).

Table \ref{table:regression_deletion}\label{footnote1} presents a summary of the logistic regression model results. As shown, we find that content posted by throwaway (anonymous) accounts is more likely to be removed by moderation than content posted by identified accounts (\textit{p} \textless{} .001). Controlling for other variables, the odds of the content posted by throwaway accounts to be removed by moderation are 71.5\% higher than identified accounts (\textit{Odds Ratio (OR)} = 1.715). We also find that content from less sensitive and less active subreddits is more likely to be removed (both \textit{p} \textless{} .001). Controlling for other variables, each one-point increase in the sensitivity of the subreddit leads to a 26.5\% decrease in the odds of its content to be removed by moderation (\textit{OR} = 0.735). Similarly, controlling for other variables, each one-point increase in the level of activity of subreddit leads to a 20.6\% decrease in the odds of its content to be removed by moderation (\textit{OR} = 0.794). Finally, we find that posts are more likely to be removed than comments (\textit{p} \textless{} .001). Controlling for other variables, the odds of posts to be removed by moderation are 87\% higher than comments (\textit{OR} = 0.130).

\begin{table*}[!htb]
    \centering
    \begin{tabular}{l|r|r|r}
         Type of moderators & \multicolumn{2}{c|}{Moderation outcome} & \\
         \hline
         & Removed (\%) & Unremoved (\%) & Total (\%)\\
         Human moderators & 8,086 (29.3\%) & 19,540 (70.7\%) & 27,626 (75.7\%)\\
         Moderation bots & 6,792 (76.4\%) & 2,096 (23.5\%) & 8,888 (24.3\%)\\
         \hline
         Total (\%) & 14,878 (40.8\%) & 21,636 (59.2\%)& 36,514 (100\%)\\
    \end{tabular}
    \caption{Human moderators vs. moderation bots on Reddit split by moderation outcome.}
    \label{table:moderation_outcome}
\end{table*}

\begin{table*}[!htb]
    \centering
    \begin{tabular}{l|rrrrr}
        & \multicolumn{5}{c}{Moderation bots vs. Human moderators}\\
        \hline
        Factors & \textit{B(SE)} & \textit{p} & \textit{2.5\% CI} & \textit{Odds Ratio} & \textit{97.5\% CI} \\
        (Intercept)  &  -0.52(0.08) & \textbf{\textless{}.001} & 0.506 & 0.592 & 0.692\\
        \begin{tabular}[c]{@{}l@{}}Throwaway accounts\\ \textit{(baseline: identified accounts)}\end{tabular} & 0.35(0.29) & .905 & 0.576 & 1.035 & 1.837\\
        \textit{Sensitivity} & -0.14(0.01) & \textbf{\textless{}.001} & 0.850 & 0.870 & 0.891\\
        Activity & 0.17(0.02) & \textbf{\textless{}.001} & 1.132 & 1.185 & 1.241\\
        \begin{tabular}[c]{@{}l@{}}Comments\\ \textit{(baseline: posts)}\end{tabular} & 0.18(0.03) & \textbf{\textless{}.001} & 1.121 & 1.201 & 1.288 \end{tabular}
    \caption{Effect of type of accounts, sensitivity of subreddit, level of activity of subreddit, entry type of content, and moderation outcome on type of moderator (moderation bots vs. human moderators) on Reddit. Human moderators were coded as 0 and moderation bots were coded as 1. Model fit: $\chi^2$ (4) = 199.31, \textit{p} \textless{} .001, Nagelkerke $R^2$ = 0.18. \textit{B} is the coefficient. \textit{SE} is the standard error. \textit{CI} is the confidence interval. We use \textit{ODDS RATIO} as the effect size.}
    \label{table:regression_moderation}
\end{table*}

\subsection{Moderation Bots vs. Human Moderators}
On Reddit, moderation logs, including those result in content removal, may be performed by either a moderation bot or a human moderator. In the previous sub-section, we explored that content posted by throwaway accounts is more likely to be removed by moderation. In this section, we explore whether the moderation bots or human moderators were responsible for such removal by moderation. As shown in Table~\ref{table:moderation_outcome}, most of moderation actions are performed by human moderators. Among all the 36,514 logs we collected, 27,626 (75.7\%) of the moderation actions were done by human moderators. As shown in Table~\ref{table:moderation_outcome}, 14,878 (40.8\%) of the moderation result in content removal. 
When human moderators perform moderation, the majority of actions (70.7\%) do not result in content removal. In contrast, moderation by bots primarily focuses on content removal, with 76.4\% of actions involving this outcome. For tasks that do not involve content removal\---constituting the majority of moderation on Reddit, there remains a significant reliance on human moderators.

To test which factors may relate to the use of moderation bots or human moderators when the content is removed, we constructed a logistic regression model using removed content only to explore the effect of the independent variables (see Section~\ref{sec:independent}) on moderation done by human moderators vs. moderation bots (dependent variable).

Table~\ref{table:regression_moderation} presents a summary of the logistic regression model results. As shown, we find that the sensitivity level of the subreddits, the active level of the subreddits, and the entry type of the content all have a significant effect (all \textit{p} \textless{} .001) on whether it will be removed by moderation bots or human moderators. The more sensitive the subreddit is, the more likely the content will be removed by human moderators (\textit{OR} = 0.870). The more active the subreddit is, the more likely the content will be removed by moderation bots (\textit{OR} = 1.185). Comparing to posts, comments are more likely to be removed by moderation bots (\textit{OR} = 1.201). However, the use of throwaway accounts does not have a significant effect (\textit{p} = .905). In other words, when posted content violated policy and needs to be removed by moderation, posting by throwaway accounts does not increase the likelihood that the moderation has to be done by human moderators.


\subsection{Fairness of the Removal}
We know from previous subsections that content is being moderated and removed on Reddit primarily by human moderators. Moreover, content posted by throwaway accounts is more likely to be removed than content posted by identified accounts. But are moderators treating content posted by throwaway accounts fairly on Reddit? To better understand RQ2 and see if moderators (human vs. bots) are treating content (anonymous vs. identified) equivalently, we measured the fairness of the content removal on Reddit. Of the 14,878 logs of the removed content, 48 (22 posts and 26 comments) came from users using throwaway accounts. Among these 48 posts or comments, 22 of them (45.8\%) were done by moderation bots, and 26 (54.2\%) were done by human moderators. Following the HCI community's norms of reliability in qualitative research (i.e., there is no need to seek agreement and inter-rater reliability due to the ease of coding and when the coding is binary; in this case, the coding was fair \textit{vs.} unfair)~\cite{mcdonald2019reliability}, the first author of this work, individually reviewed these 48 logs (Group A). Since content posted on Reddit could be removed due to both the violation of the policy of Reddit and the violation of the rules of each subreddit, the first author first reviewed the content policy of Reddit and then reviewed the rules of each subreddit that the 48 logs belong to. The first author then manually reviewed each removed post or comment and determined if the post or comment violated any policy or rule and should be removed. Although the coding process was relatively easy and straightforward, we confirmed our coding with two research assistants who are both heavy Reddit users. We reached agreement without the need to have a discussion. 

Note that subreddit rules do differ across Reddit, but we didn't check rules across subreddits. Using the same approach illustrated above, the first author also reviewed 48 logs per group from the following three groups: removed content posted by identified accounts (Group B), unremoved content posted by throwaway accounts (Group C), and unremoved content posted by identified accounts (Group D). Each group of 48 logs was randomly selected based on the subreddits from Group A (Logs of groups are from 24 subreddits). For example, if the first subreddit in Group A has two logs, then we randomly selected two logs from Group B, C, and D from the same subreddit, respectively. We iterated the process until we selected 48 logs for Group B, C, and D, respectively.

\begin{table*}[!htb]
\begin{tabular}{c|c|c|l}
Group & Throwaway? & Removed? & Content\\
\multirow{2}{*}{A} & \multirow{2}{*}{Yes} & \multirow{2}{*}{Yes} & Join this discord server for free daily hot pics and vids and content from those premium snapchat people:\\
 &  &  & https://discord.gg/xxxxx  \\
B & No & Yes & White people problems\\
C & Yes & No & Wow thanks! We need to do more research on this Atlantic council as we never heard of that.\\
D & No & No & What's your prediction on BTC for the next 10 years?\\
\end{tabular}
\caption{Examples of removed and unremoved content, split by identity. Identifiable information are removed.}
\label{tab:mod_example}
\end{table*}

We found that posts or comments from Group A and Group B indeed violate either Reddit's content policy or the subreddit's rule and thus should have been removed. We also found that posts or comments from Group C and D did not violate any policy or rule and should have been kept. Table~\ref{tab:mod_example} provides example posts/comments from each group. To sum up, we found that content posted by throwaway accounts is treated the same as content posted by identified accounts.
\section{Discussion}
In this paper, we explore the use of anonymity in an often forgotten and inaccessible area of the content on SMPs\---the moderated content. Our results illustrate that although the use of throwaway accounts does not increase the likelihood of content being removed by human moderators, content posted by throwaway accounts is more likely to be removed by moderation, and the moderators' removal decisions are usually fair. Our findings provide two areas of insight for SMPs practitioners: anonymous content and the content moderation mechanisms both of which we discuss in detail below. 

\subsection{Moderation as a Solution for Protecting Privacy and Ensuring Content Quality on SMPs}
\subsubsection{People choose to be anonymous on SMPs especially for highly sensitive topics}
People want to be anonymous online, including on SMPs, for many reasons. First, people's past experiences and life situations lead them to choose to be anonymous online~\cite{kang2013people}. In the same study, 93\% of the participants have the experience of being anonymous for social activities, and 57\% have participated in special interest groups anonymously. Being anonymous online, especially when interacting with highly sensitive content (e.g., highly sensitive subreddits on Reddit), may provide people more control over personal information disclosure, protect their personal safety, provide more emotional benefits, and make them feel free to express opinions~\cite{kang2013people}. On Reddit, people may use throwaway accounts, especially in highly sensitive subreddits, to look for information or social support. For example, in the subreddits related to sex abuse on Reddit, ``anonymous commenting enables commenters to share intimate content such as reciprocal disclosures and supportive messages''~\cite{andalibi2018social}. People also feel more comfortable disclosing highly sensitive information (e.g., pregnancy loss) on identified SMPs (e.g., Facebook) after participation anonymously on Reddit~\cite{andalibi2018announcing}. SMPs like Reddit that allow throwaway accounts should make it easy for users to switch between identities without logging in and out. SMPs like Reddit may also consider using other approaches to let users engage anonymously more easily. Currently, Reddit users have to use some sort of a workaround (e.g., creating throwaway accounts) to stay anonymous. This might discourage or prevent users from engaging with SMPs. For example, people who are less knowledgeable about SMPs may not be aware of how to create a throwaway account. Prior work shows that people who have concerns about how social media can adversely affect their relationships with others will use SMPs less frequently~\cite{page2019communication}. The use of throwaway accounts/anonymous accounts could reduce these concerns.

\subsubsection{Content posted by throwaway accounts is more likely to be removed by moderation}
On the other hand, there are also downsides to allowing anonymity online. Friedman and Resnick~\cite{friedman2001social} highlighted the issue of ``cheap pseudonyms'', where the ease of creating new online identities with minimal effort or cost can encourage misbehavior without the risk of harming one's reputation. This issue may be even more prevalent on SMPs, where obtaining new identities often requires just a few simple registration steps. For instance, studies have shown that removing the option for anonymous participation can improve the quality of comments on online news sites~\cite{fredheim2015anonymity} and online communities of practice~\cite{kilner2005anonymity}. 

Similarly, in our study, we find that content posted by throwaway accounts is more likely to be removed. The online disinhibition theory describes a phenomenon where people may have a lack of restraint on the Internet, which may lead to toxic behavior~\cite{suler2004online}. In this theory, anonymity is one of the factors that could lead to toxic disinhibition. Similarly, in the social identity theory of deindividuation~\cite{reicher1995social}, anonymity is one of the factors that could lead people to lose self-control over behavior in groups (e.g., SMPs like Reddit). Based on these theories and our empirical study, moderators need to exercise heightened vigilance toward throwaway accounts or anonymous users, as theoretically, people may use throwaway accounts or anonymous identities deliberately to misbehave or troll on SMPs. However, the solution is not to disallow throwaway accounts and anonymous identity on SMPs as this would make it very difficult for people who have privacy needs and want to seek information online, especially around highly sensitive topics. While this study does not definitively resolve the ongoing debate regarding throwaway accounts and anonymous identities, we propose that a robust and effectively implemented moderation mechanism may offer a viable solution. In an ideal scenario, moderation can allow users with privacy needs to participate anonymously while ensuring the timely removal of harmful content and dissuade users who deliberately misuse anonymity.

\subsection{SMPs Should Reduce the Cost to and of Human Moderators}
\subsubsection{Human moderation on Reddit is generally fair and is heavily relied upon for enforcing platform rules and community standards}
Many users do not perceive the removal of their content by moderation as justified~\cite{jhaver2019did}. Additionally, aside from publicmodlogs, Reddit's moderation system lacks transparency~\cite{juneja2020through}. However, in our study, we find that content removal decisions made by human moderators on Reddit are generally fair. Unlike the moderation bots, which have pre-defined rules and strictly enforce them, the moderation work by human moderators is more flexible. Although different moderators may take different moderation actions, especially in grey areas, human moderators are well aware of their tasks' subjective nature~\cite{diakopoulos2011towards}. Our results reveal that most of the moderation actions are still done by human moderators on Reddit, especially for the moderation actions that are not just removing a post or a comment.

\subsubsection{The cost to and of human moderators is high}
Human moderation has human costs~\cite{humancost:1,humancost:2}, including that it is time-consuming and takes a psychological toll. For example, 40\% of the participants (journalists and human rights workers) from a survey study reported that ``viewing distressing eyewitness media has had a negative impact on their personal lives''~\cite{dubberley2015making}. As a human moderator, someone may spend hours each day viewing those distressing eyewitness media to support their online communities. Moreover, on Reddit, human moderators all voluntarily take on moderation tasks. On other SMPs such as Facebook, human moderators are paid, but the compensation is low and does not stop human moderators from presenting with PTSD-like symptoms~\cite{Facebook}. All of these could lead to the grievances of human moderators: ``the work of moderators and the precarity of their position with company policies''~\cite{matias2016going}. The grievances of human moderators may cause other further issues, such as the Reddit Blackout of July 2015~\cite{matias2015just}. During the Reddit Blackout of July 2015, human moderators of NSFW (highly sensitive) subreddits were more likely to blackout than human moderators of SFW (less sensitive) subreddits~\cite{matias2016going}. Human moderators of highly sensitive subreddits are probably seeing more distressing eyewitness media than human moderators of less sensitive subreddits, which may cause more human costs such as more severe PTSD-like symptoms or grievances.

Another aspect of cost is that doing more moderation work may increase human moderators' bias regarding the content they encounter. For example, trolling, which ``includes flaming, griefing, swearing, or personal attacks, including behavior outside the acceptable bounds defined by several community guidelines for discussion forums''~\cite{cheng2017anyone}, is a common behavior on SMPs including Reddit~\cite{merritt2012analysis}. Trolling, by definition, violates the community norms on Reddit~\cite{chandrasekharan2018internet}, will very likely to be removed by moderation. 
Human moderators, who see a lot of trolling posts on a regular basis, may also be susceptible. Seeing trolling posts could cause their own misuse of authority (e.g., retaliation), although this seems dependent on personality traits~\cite{skarlicki1999personality}. This could potentially increase the bias of human moderation.

\subsubsection{New forms of moderation are needed to reduce human cost}
Volunteer governance continues to be the common approach to managing social relations (e.g., human moderation) in online communities~\cite{matias2016civic}. Besides the positive effects of shielding a community from undesirable content, the content removal can also help to improve the comment's author's future behavior~\cite{srinivasan2019content}. However, as we discussed above, human moderation has human costs. Alternative moderation approaches such as the moderation bots on Reddit could participate more in content moderation to reduce the dependency on human moderation. Currently, on Reddit, the most popular moderation bot - AutoModerator, can only ``remove or flair posts by domain or keyword''.\footnote{https://support.reddithelp.com/hc/articles/15484574206484} Although we find that more active subreddits are more likely to have moderation bots remove content, the human labor of moderation is necessary~\cite{gillespie2018custodians}. It is challenging to completely replace human moderators with moderation bots due to the complexity of language and community norms. Especially for grey areas, human moderators might be more flexible than moderation bots. In fact, we find that more sensitive subreddits are more likely to have human moderators remove content. However, we argue that SMPs should take better care of human moderators and develop effective tools to support human moderators for content moderation. For example, an AI-backed sociotechnical moderation system can already detect close to 90\% of the comments that would be removed by human moderators~\cite{chandrasekharan2019crossmod}, which may also help with the gray areas. It can significantly reduce the costs to and of human moderators. But, humans need to remain in the loop to ensure moderation bots are helping and not removing content that human moderators would want to remain.

\subsection{Implications for Moderation on SMPs}
While our study finds that moderation by human moderators on Reddit is generally fair, having posts moderated even by humans is not always well-received by Reddit users~\cite{jhaver2019did}. Previous research highlights the lack of transparency in Reddit's moderation processes~\cite{juneja2020through}. Increasing moderation transparency, such as providing explanations for content removal, has been shown to reduce the likelihood of users violating rules in the future~\cite{jhaver2019does}. We argue that enhancing transparency could also improve users' perceptions of moderation fairness, which may ultimately lower the burden on human moderators and reduce moderation costs over time. Future research could investigate which aspects of the moderation process are most critical to improving users' sense of fairness and how moderation systems can efficiently deliver these features for moderators to implement.

Future moderation systems on SMPs should incorporate context sensitivity in their design. Future research could explore how to enhance these systems to classify posts and comments based on sensitivity levels, enabling moderators to identify and prioritize monitoring of sensitive content more effectively. For SMPs that cover a broad range of topics (e.g., some subreddits address more sensitive issues than others), practitioners should consider reallocating moderation resources to prioritize the moderation of highly sensitive topics.
\section{Limitations}
Our study has several limitations. Firstly, our research scope was constrained by the availability of publicmodlogs. While we assessed subreddit sensitivity and activity levels across a diverse range, there are more active and diverse subreddits from which we could not ethically/legally collect data. Therefore, our findings may not be generalizable to the entire Reddit community or other SMPs. Future research could expand upon our work if Reddit would provide APIs for its moderation logs to the research community. Secondly, while we identified moderation bots and throwaway accounts using established methods and manual verification by researchers, there may be false negatives in our data. For example, despite the community norm of self-declaring throwaway accounts~\cite{gagnon2013disinhibition}, some users may not adhere to this practice, leading to potential omissions in our analysis. However, we believe that by following literature and best practices, the number of missed throwaway accounts should be minimal.

Identifying false negative throwaway accounts is challenging. Analyzing user activity and account history can help detect some overlooked throwaway accounts, but this approach also risks producing false positives~\cite{garg2021using}. For instance, accounts with only a single post or accounts affected by customer churn may be mistakenly labeled as throwaway accounts using this method. Future research could develop more automated and accurate approaches to identify moderation bots and throwaway accounts on Reddit. Thirdly, while transparency is a primary reason for subreddits to make their moderation logs public, moderators also believe this practice increases accountability~\cite{juneja2020through}. Therefore, moderators of subreddits participating in public moderation logs may behave differently than those in non-participating subreddits. Future empirical studies are needed to explore these potential differences.
\section{Conclusion}
SMP users often have privacy concerns that motivate them to participate anonymously, such as through the use of throwaway accounts. Content posted on SMPs is typically moderated by both human moderators and moderation bots. This paper utilizes publicmodlogs to analyze throwaway accounts and moderation actions on Reddit. Our findings reveal that while the use of throwaway accounts does not increase the likelihood of content being removed by human moderators, content posted by throwaway accounts is generally more likely to violate rules and is more likely to be removed by moderation. Additionally, we find that human moderators continue to perform the majority of moderation tasks on Reddit. We highlight the need to use moderation as a solution for protecting privacy and ensuring content quality on SMPs. We propose that SMP practitioners should also seek to reduce the burden on human moderators through the development of more efficient and effective moderation tools and strategies.

\section*{Acknowledgment}
We thank anonymous reviewers for their insightful feedback on this work. The authors also thank John Xue and Keith Stoddard for their assistance on labeling data. Finally, we thank Dr. Bart Knijnenburg and Dr. Emily Sidnam-Mauch for their comments on early versions of this work.



%
\bibliographystyle{IEEEtran}
\bibliography{ref.bib}

\begin{thebibliography}{10}
\providecommand{\url}[1]{#1}
\csname url@samestyle\endcsname
\providecommand{\newblock}{\relax}
\providecommand{\bibinfo}[2]{#2}
\providecommand{\BIBentrySTDinterwordspacing}{\spaceskip=0pt\relax}
\providecommand{\BIBentryALTinterwordstretchfactor}{4}
\providecommand{\BIBentryALTinterwordspacing}{\spaceskip=\fontdimen2\font plus
\BIBentryALTinterwordstretchfactor\fontdimen3\font minus \fontdimen4\font\relax}
\providecommand{\BIBforeignlanguage}[2]{{%
\expandafter\ifx\csname l@#1\endcsname\relax
\typeout{** WARNING: IEEEtran.bst: No hyphenation pattern has been}%
\typeout{** loaded for the language `#1'. Using the pattern for}%
\typeout{** the default language instead.}%
\else
\language=\csname l@#1\endcsname
\fi
#2}}
\providecommand{\BIBdecl}{\relax}
\BIBdecl

\bibitem{leavitt2015throwaway}
A.~Leavitt, ``This is a throwaway account: Temporary technical identities and perceptions of anonymity in a massive online community,'' in \emph{Proceedings of the 18th ACM Conference on Computer Supported Cooperative Work \& Social Computing}.\hskip 1em plus 0.5em minus 0.4em\relax ACM, 2015, pp. 317--327.

\bibitem{rainie2013anonymity}
L.~Rainie, S.~Kiesler, R.~Kang, M.~Madden, M.~Duggan, S.~Brown, and L.~Dabbish, ``Anonymity, privacy, and security online,'' \emph{Pew Research Center}, vol.~5, 2013.

\bibitem{andalibi2016understanding}
N.~Andalibi, O.~L. Haimson, M.~De~Choudhury, and A.~Forte, ``Understanding social media disclosures of sexual abuse through the lenses of support seeking and anonymity,'' in \emph{Proceedings of the 2016 CHI Conference on Human Factors in Computing Systems}.\hskip 1em plus 0.5em minus 0.4em\relax ACM, 2016, pp. 3906--3918.

\bibitem{andalibi2018social}
N.~Andalibi, O.~L. Haimson, M.~D. Choudhury, and A.~Forte, ``Social support, reciprocity, and anonymity in responses to sexual abuse disclosures on social media,'' \emph{ACM Transactions on Computer-Human Interaction (TOCHI)}, vol.~25, no.~5, p.~28, 2018.

\bibitem{kang2013people}
R.~Kang, S.~Brown, and S.~Kiesler, ``Why do people seek anonymity on the internet?: informing policy and design,'' in \emph{Proceedings of the SIGCHI Conference on Human Factors in Computing Systems}.\hskip 1em plus 0.5em minus 0.4em\relax ACM, 2013, pp. 2657--2666.

\bibitem{cheng2017anyone}
J.~Cheng, M.~Bernstein, C.~Danescu-Niculescu-Mizil, and J.~Leskovec, ``Anyone can become a troll: Causes of trolling behavior in online discussions,'' in \emph{Proceedings of the 2017 ACM Conference on Computer Supported Cooperative Work and Social Computing}.\hskip 1em plus 0.5em minus 0.4em\relax ACM, 2017, pp. 1217--1230.

\bibitem{kraut2012building}
R.~E. Kraut and P.~Resnick, \emph{Building successful online communities: Evidence-based social design}.\hskip 1em plus 0.5em minus 0.4em\relax Mit Press, 2012.

\bibitem{herring2002searching}
S.~Herring, K.~Job-Sluder, R.~Scheckler, and S.~Barab, ``Searching for safety online: Managing" trolling" in a feminist forum,'' \emph{The information society}, vol.~18, no.~5, pp. 371--384, 2002.

\bibitem{grimmelmann2015virtues}
J.~Grimmelmann, ``The virtues of moderation,'' \emph{Yale JL \& Tech.}, vol.~17, p.~42, 2015.

\bibitem{guo2020anonymity}
C.~Guo and K.~Caine, ``Anonymity in questions and answers about health,'' in \emph{Proceedings of the Human Factors and Ergonomics Society Annual Meeting}, vol.~64, no.~1.\hskip 1em plus 0.5em minus 0.4em\relax SAGE Publications Sage CA: Los Angeles, CA, 2020, pp. 658--662.

\bibitem{ponzanelli2014improving}
L.~Ponzanelli, A.~Mocci, A.~Bacchelli, M.~Lanza, and D.~Fullerton, ``Improving low quality stack overflow post detection,'' in \emph{2014 IEEE international conference on software maintenance and evolution}.\hskip 1em plus 0.5em minus 0.4em\relax IEEE, 2014, pp. 541--544.

\bibitem{guo2021anonymity}
C.~Guo and K.~Caine, ``Anonymity, user engagement, quality, and trolling on q\&amp;a sites,'' \emph{Proceedings of the ACM on Human-Computer Interaction}, vol.~5, no. CSCW1, pp. 1--27, 2021.

\bibitem{kiene2019technological}
C.~Kiene, J.~A. Jiang, and B.~M. Hill, ``Technological frames and user innovation: Exploring technological change in community moderation teams,'' \emph{Proceedings of the ACM on Human-Computer Interaction}, vol.~3, no. CSCW, pp. 1--23, 2019.

\bibitem{jhaver2019human}
S.~Jhaver, I.~Birman, E.~Gilbert, and A.~Bruckman, ``Human-machine collaboration for content regulation: The case of reddit automoderator,'' \emph{ACM Transactions on Computer-Human Interaction (TOCHI)}, vol.~26, no.~5, pp. 1--35, 2019.

\bibitem{humancost:1}
B.~Powers, ``The human cost of monitoring the internet,'' 2017, retrieved September 15, 2024 from https://www.rollingstone.com/culture/culture-features/the-human-cost-of-monitoring-the-internet-202291/.

\bibitem{humancost:2}
A.~Arsht and D.~Etcovitch, ``The human cost of online content moderation,'' 2018, retrieved September 15, 2024 from https://jolt.law.harvard.edu/digest/the-human-cost-of-online-content-moderation.

\bibitem{haimson2016constructing}
O.~L. Haimson and A.~L. Hoffmann, ``Constructing and enforcing ``authentic'' identity online: Facebook, real names, and non-normative identities,'' \emph{First Monday}, vol.~21, no.~6, 2016.

\bibitem{facebook:fake}
A.~Schultz, ``How does facebook measure fake accounts?'' 2019, retrieved September 15, 2024 from https://about.fb.com/news/2019/05/fake-accounts/.

\bibitem{peddinti2014cloak}
S.~T. Peddinti, A.~Korolova, E.~Bursztein, and G.~Sampemane, ``Cloak and swagger: Understanding data sensitivity through the lens of user anonymity,'' in \emph{2014 IEEE Symposium on Security and Privacy}.\hskip 1em plus 0.5em minus 0.4em\relax IEEE, 2014, pp. 493--508.

\bibitem{gagnon2013disinhibition}
T.~Gagnon, ``The disinhibition of reddit users,'' \emph{Adele Richardson's Spring}, 2013.

\bibitem{marx1999s}
G.~T. Marx, ``What's in a name? some reflections on the sociology of anonymity,'' \emph{The Information Society}, vol.~15, no.~2, pp. 99--112, 1999.

\bibitem{marx2019windows}
G.~Marx, \emph{Windows into the soul: Surveillance and society in an age of high technology}.\hskip 1em plus 0.5em minus 0.4em\relax University of Chicago Press, 2019.

\bibitem{marx2004internet}
G.~T. Marx, ``Internet anonymity as a reflection of broader issues involving technology and society,'' \emph{Asia-Pacific Review}, vol.~11, no.~1, pp. 142--166, 2004.

\bibitem{urbanski2013upvoting}
D.~Urbanski, ``Upvoting the audience: a burkean analysis of reddit,'' 2013.

\bibitem{pavalanathan2015identity}
U.~Pavalanathan and M.~De~Choudhury, ``Identity management and mental health discourse in social media,'' in \emph{Proceedings of the 24th International Conference on World Wide Web}.\hskip 1em plus 0.5em minus 0.4em\relax ACM, 2015, pp. 315--321.

\bibitem{dosono2017exploring}
B.~Dosono, B.~Semaan, and J.~Hemsley, ``Exploring aapi identity online: Political ideology as a factor affecting identity work on reddit,'' in \emph{Proceedings of the 2017 CHI Conference Extended Abstracts on Human Factors in Computing Systems}.\hskip 1em plus 0.5em minus 0.4em\relax ACM, 2017, pp. 2528--2535.

\bibitem{de2014mental}
M.~De~Choudhury and S.~De, ``Mental health discourse on reddit: Self-disclosure, social support, and anonymity,'' in \emph{Eighth International AAAI Conference on Weblogs and Social Media}, 2014.

\bibitem{ammari2019self}
T.~Ammari, S.~Schoenebeck, and D.~Romero, ``Self-declared throwaway accounts on reddit: How platform affordances and shared norms enable parenting disclosure and support,'' \emph{Proceedings of the ACM on Human-Computer Interaction}, vol.~3, no. CSCW, pp. 1--30, 2019.

\bibitem{johnson2000anonymity}
D.~Johnson, ``Anonymity and the internet,'' \emph{The Futurist}, vol.~34, no.~4, p.~12, 2000.

\bibitem{krappitz2012troll}
S.~Krappitz, ``Troll culture,'' \emph{Retrieved September 15, 2024 from http://wwwwwwwww.at/downloads/troll-culture.pdf}, 2012.

\bibitem{reicher1995social}
S.~D. Reicher, R.~Spears, and T.~Postmes, ``A social identity model of deindividuation phenomena,'' \emph{European review of social psychology}, vol.~6, no.~1, pp. 161--198, 1995.

\bibitem{chandrasekharan2018internet}
E.~Chandrasekharan, M.~Samory, S.~Jhaver, H.~Charvat, A.~Bruckman, C.~Lampe, J.~Eisenstein, and E.~Gilbert, ``The internet's hidden rules: An empirical study of reddit norm violations at micro, meso, and macro scales,'' \emph{Proceedings of the ACM on Human-Computer Interaction}, vol.~2, no. CSCW, p.~32, 2018.

\bibitem{tran2020anonymity}
C.~Tran, K.~Champion, A.~Forte, B.~M. Hill, and R.~Greenstadt, ``Are anonymity-seekers just like everybody else? an analysis of contributions to wikipedia from tor,'' in \emph{2020 IEEE Symposium on Security and Privacy (SP)}.\hskip 1em plus 0.5em minus 0.4em\relax IEEE, 2020, pp. 186--202.

\bibitem{bogers2014social}
T.~Bogers and R.~N. Wernersen, ``How'social'are social news sites? exploring the motivations for using reddit. com,'' in \emph{Proceedings of the iConference 2014}.\hskip 1em plus 0.5em minus 0.4em\relax iSchools, 2014, pp. 329--344.

\bibitem{moore2017redditors}
C.~Moore and L.~Chuang, ``Redditors revealed: Motivational factors of the reddit community,'' in \emph{Proceedings of the 50th Hawaii International Conference on System Sciences}, 2017.

\bibitem{wang2013wisdom}
G.~Wang, K.~Gill, M.~Mohanlal, H.~Zheng, and B.~Y. Zhao, ``Wisdom in the social crowd: an analysis of quora,'' in \emph{Proceedings of the 22nd International Conference on World Wide Web}.\hskip 1em plus 0.5em minus 0.4em\relax ACM, 2013, pp. 1341--1352.

\bibitem{li2015answer}
L.~Li, D.~He, W.~Jeng, S.~Goodwin, and C.~Zhang, ``Answer quality characteristics and prediction on an academic q\&a site: A case study on researchgate,'' in \emph{Proceedings of the 24th International Conference on World Wide Web}.\hskip 1em plus 0.5em minus 0.4em\relax ACM, 2015, pp. 1453--1458.

\bibitem{kiel2017could}
J.~V. Kiel~Long, S.~Sutton, P.~Brooker, T.~Feltwell, B.~Kirman, J.~Barnett, and S.~Lawson, ``Could you define that in bot terms?: Requesting, creating and using bots on reddit,'' in \emph{Proceedings of the 2017 CHI Conference on Human Factors in Computing Systems. ACM}, 2017, pp. 3488--3500.

\bibitem{AutoModerator}
Reddit, ``Automoderator,'' 2023, retrieved September 15, 2024 from https://mods.reddithelp.com/hc/en-us/articles/360002561632-AutoModerator.

\bibitem{teblunthuis2021effects}
N.~TeBlunthuis, B.~M. Hill, and A.~Halfaker, ``Effects of algorithmic flagging on fairness: Quasi-experimental evidence from wikipedia,'' \emph{Proceedings of the ACM on Human-Computer Interaction}, vol.~5, no. CSCW1, pp. 1--27, 2021.

\bibitem{juneja2020through}
P.~Juneja, D.~Rama~Subramanian, and T.~Mitra, ``Through the looking glass: Study of transparency in reddit's moderation practices,'' \emph{Proceedings of the ACM on Human-Computer Interaction}, vol.~4, no. GROUP, pp. 1--35, 2020.

\bibitem{li2022all}
H.~Li, B.~Hecht, and S.~Chancellor, ``All that’s happening behind the scenes: Putting the spotlight on volunteer moderator labor in reddit,'' in \emph{Proceedings of the International AAAI Conference on Web and Social Media}, vol.~16, 2022, pp. 584--595.

\bibitem{li2022measuring}
------, ``Measuring the monetary value of online volunteer work,'' in \emph{Proceedings of the International AAAI Conference on Web and Social Media}, vol.~16, 2022, pp. 596--606.

\bibitem{van2013faceless}
E.~Van~der Nagel, ``Faceless bodies: Negotiating technological and cultural codes on reddit gonewild,'' \emph{Scan: Journal of Media Arts Culture}, vol.~10, no.~2, pp. 1--10, 2013.

\bibitem{lee1990problems}
R.~M. Lee and C.~M. Renzetti, ``The problems of researching sensitive topics: An overview and introduction,'' 1990.

\bibitem{hallgren2012computing}
K.~A. Hallgren, ``Computing inter-rater reliability for observational data: an overview and tutorial,'' \emph{Tutorials in Quantitative Methods for Psychology}, vol.~8, no.~1, p.~23, 2012.

\bibitem{bland1997statistics}
J.~M. Bland and D.~G. Altman, ``Statistics notes: Cronbach's alpha,'' \emph{Bmj}, vol. 314, no. 7080, p. 572, 1997.

\bibitem{jhaver2019does}
S.~Jhaver, A.~Bruckman, and E.~Gilbert, ``Does transparency in moderation really matter? user behavior after content removal explanations on reddit,'' \emph{Proceedings of the ACM on Human-Computer Interaction}, vol.~3, no. CSCW, pp. 1--27, 2019.

\bibitem{zimmer2010but}
M.~Zimmer, ````but the data is already public'': on the ethics of research in facebook,'' \emph{Ethics and information technology}, vol.~12, no.~4, pp. 313--325, 2010.

\bibitem{fisher2010terms}
D.~Fisher, D.~W. Mcdonald, A.~L. Brooks, and E.~F. Churchill, ``Terms of service, ethics, and bias: Tapping the social web for cscw research,'' \emph{Computer Supported Cooperative Work (CSCW), Panel discussion}, 2010.

\bibitem{fiesler2015ethics}
C.~Fiesler, A.~Young, T.~Peyton, A.~S. Bruckman, M.~Gray, J.~Hancock, and W.~Lutters, ``Ethics for studying online sociotechnical systems in a big data world,'' in \emph{Proceedings of the 18th ACM Conference Companion on Computer Supported Cooperative Work \& Social Computing}, 2015, pp. 289--292.

\bibitem{munteanu2019sigchi}
C.~Munteanu, A.~Bruckman, M.~Muller, C.~Frauenberger, C.~Fiesler, R.~E. Kraut, K.~Shilton, and J.~Waycott, ``Sigchi research ethics town hall,'' in \emph{Extended Abstracts of the 2019 CHI Conference on Human Factors in Computing Systems}, 2019, pp. 1--6.

\bibitem{bruckman2017cscw}
A.~S. Bruckman, C.~Fiesler, J.~Hancock, and C.~Munteanu, ``Cscw research ethics town hall: Working towards community norms,'' in \emph{Companion of the 2017 ACM Conference on Computer Supported Cooperative Work and Social Computing}, 2017, pp. 113--115.

\bibitem{o2007caution}
R.~M. O’brien, ``A caution regarding rules of thumb for variance inflation factors,'' \emph{Quality \& quantity}, vol.~41, no.~5, pp. 673--690, 2007.

\bibitem{mcdonald2019reliability}
N.~McDonald, S.~Schoenebeck, and A.~Forte, ``Reliability and inter-rater reliability in qualitative research: Norms and guidelines for cscw and hci practice,'' \emph{Proceedings of the ACM on Human-Computer Interaction}, vol.~3, no. CSCW, pp. 1--23, 2019.

\bibitem{andalibi2018announcing}
N.~Andalibi and A.~Forte, ``Announcing pregnancy loss on facebook: A decision-making framework for stigmatized disclosures on identified social network sites,'' in \emph{Proceedings of the 2018 CHI Conference on Human Factors in Computing Systems}.\hskip 1em plus 0.5em minus 0.4em\relax ACM, 2018, p. 158.

\bibitem{page2019communication}
X.~Page, R.~G. Anaraky, and B.~P. Knijnenburg, ``How communication style shapes relationship boundary regulation and social media adoption,'' in \emph{Proceedings of the 10th International Conference on Social Media and Society}, 2019, pp. 126--135.

\bibitem{friedman2001social}
E.~J. Friedman and P.~Resnick, ``The social cost of cheap pseudonyms,'' \emph{Journal of Economics \& Management Strategy}, vol.~10, no.~2, pp. 173--199, 2001.

\bibitem{fredheim2015anonymity}
R.~Fredheim, A.~Moore, and J.~Naughton, ``Anonymity and online commenting: The broken windows effect and the end of drive-by commenting,'' in \emph{Proceedings of the ACM Web Science Conference}.\hskip 1em plus 0.5em minus 0.4em\relax ACM, 2015, p.~11.

\bibitem{kilner2005anonymity}
P.~G. Kilner and C.~M. Hoadley, ``Anonymity options and professional participation in an online community of practice,'' in \emph{Proceedings of th 2005 Conference on Computer Support for Collaborative Learning}.\hskip 1em plus 0.5em minus 0.4em\relax International Society of the Learning Sciences, 2005, pp. 272--280.

\bibitem{suler2004online}
J.~Suler, ``The online disinhibition effect,'' \emph{Cyberpsychology \& behavior}, vol.~7, no.~3, pp. 321--326, 2004.

\bibitem{jhaver2019did}
S.~Jhaver, D.~S. Appling, E.~Gilbert, and A.~Bruckman, ````did you suspect the post would be removed?'' understanding user reactions to content removals on reddit,'' \emph{Proceedings of the ACM on Human-Computer Interaction}, vol.~3, no. CSCW, pp. 1--33, 2019.

\bibitem{diakopoulos2011towards}
N.~Diakopoulos and M.~Naaman, ``Towards quality discourse in online news comments,'' in \emph{Proceedings of the ACM 2011 Conference on Computer Supported Cooperative Work}.\hskip 1em plus 0.5em minus 0.4em\relax ACM, 2011, pp. 133--142.

\bibitem{dubberley2015making}
S.~Dubberley, E.~Griffin, and H.~M. Bal, ``Making secondary trauma a primary issue: A study of eyewitness media and vicarious trauma on the digital frontline,'' \emph{Istanbul: Eyewitness Media Hub}, 2015.

\bibitem{Facebook}
B.~Feldman, ``Facebook can’t solve its problems by throwing bodies at them,'' 2019, retrieved September 15, 2024 from http://nymag.com/intelligencer/2019/02/facebooks-content-moderators-are-low-paid-developing-ptsd.html.

\bibitem{matias2016going}
J.~N. Matias, ``Going dark: Social factors in collective action against platform operators in the reddit blackout,'' in \emph{Proceedings of the 2016 CHI Conference on Human Factors in Computing Systems}.\hskip 1em plus 0.5em minus 0.4em\relax ACM, 2016, pp. 1138--1151.

\bibitem{matias2015just}
J.~Matias, ``What just happened on reddit? understanding the moderator blackout,'' \emph{Social Media Collective, Microsoft England}, 2015.

\bibitem{merritt2012analysis}
E.~Merritt, ``An analysis of the discourse of internet trolling: A case study of reddit. com,'' Ph.D. dissertation, 2012.

\bibitem{skarlicki1999personality}
D.~P. Skarlicki, R.~Folger, and P.~Tesluk, ``Personality as a moderator in the relationship between fairness and retaliation,'' \emph{Academy of Management Journal}, vol.~42, no.~1, pp. 100--108, 1999.

\bibitem{matias2016civic}
J.~N. Matias, ``The civic labor of online moderators,'' in \emph{Internet Politics and Policy conference, Oxford, United Kingdom}, 2016.

\bibitem{srinivasan2019content}
K.~B. Srinivasan, C.~Danescu-Niculescu-Mizil, L.~Lee, and C.~Tan, ``Content removal as a moderation strategy: Compliance and other outcomes in the changemyview community,'' \emph{Proceedings of the ACM on Human-Computer Interaction}, vol.~3, no. CSCW, pp. 1--21, 2019.

\bibitem{gillespie2018custodians}
T.~Gillespie, \emph{Custodians of the Internet: Platforms, content moderation, and the hidden decisions that shape social media}.\hskip 1em plus 0.5em minus 0.4em\relax Yale University Press, 2018.

\bibitem{chandrasekharan2019crossmod}
E.~Chandrasekharan, C.~Gandhi, M.~W. Mustelier, and E.~Gilbert, ``Crossmod: A cross-community learning-based system to assist reddit moderators,'' \emph{Proceedings of the ACM on Human-Computer Interaction}, vol.~3, no. CSCW, pp. 1--30, 2019.

\bibitem{garg2021using}
R.~Garg, Y.~Kapadia, and S.~Sengupta, ``Using the lenses of emotion and support to understand unemployment discourse on reddit,'' \emph{Proceedings of the ACM on Human-Computer Interaction}, vol.~5, no. CSCW1, pp. 1--24, 2021.

\end{thebibliography}

\section{Moderation actions collected via publicmodlogs}
\begin{table}[!htb]
    \centering
    \begin{tabular}{l|r}
         Moderation actions & N\\
         \hline
         Accept moderation invite & 60\\
         Add community topics & 2\\
         Add contributor & 24\\
         Approve comment & 3,374\\
         Approve link & 5,538\\
         Ban user & 902\\
         Collections & 1\\
         Community styling & 52\\
         Community widgets & 335\\
         Create rule & 21\\
         Delete rule & 22\\
         Distinguish & 1,982\\
         Edit flair & 2,962\\
         Edit rule & 24\\
         Edit settings & 124\\
         Ignore reports & 549\\
         Invite moderator & 63\\
         Lock & 434\\
         Mark NSFW & 26\\
         Mark original content & 51\\
         Modmail enrollment & 1\\
         Mute user & 75\\
         \textbf{Remove comment} & 5,685\\
         Remove contributor & 4\\
         \textbf{Remove link} & 6,651\\
         Remove moderator & 35\\
         Set contest mode & 16\\
         Set permissions & 9\\
         Set suggested sort & 6\\
         \textbf{Spam comment} & 278\\
         \textbf{Spam link} & 2,264\\
         Spoiler & 9\\
         Sticky & 1,988\\
         Unban user & 141\\
         Unignore reports & 92\\
         Uninvite moderator & 5\\
         Unlock & 29\\
         Unmute user & 72\\
         Unset contest mode & 12\\
         Unspoiler & 3\\
         Unsticky & 836\\
         Wiki page listed & 6\\
         Wiki revise & 1,751\\
    \end{tabular}
    \caption{Moderation logs collected via publicmodlogs on Reddit, logs that result in content removal are highlighted in bold}
    \label{table:moderation_full}
\end{table}


\end{document}